# PLASMA PROCESSES AT THE SCALES OF GALAXIES AND CLUSTERS OF GALAXIES

PRATEEK SHARMA*

*Matter on large scales in the Universe is distributed in form of sheets, filaments and spherical halos. Most of the known matter in the Universe is in a diffuse plasma state — in halos around galaxies and clusters of galaxies, and in the intergalactic medium tracing the large scale cosmological filaments. Recent multi-wavelength observations have informed us about the state of this dilute plasma that covers a range of densities and temperatures. Progress in high resolution numerical simulations allows us to understand this diffuse plasma, but there are several puzzles that involve a complex interplay of processes such as cooling, feedback heating, gravity, and turbulence.[†]*

## Introduction

Most readers may have encountered the statement "99% of the Universe is in the plasma state." This statement needs to be qualified. The modern budgeting of mass/energy (recall $E = mc^2$, that is mass and energy are equivalent) in the Universe has revealed that the normal matter, described by the celebrated standard model of physics, comprises only 5% to the total mass/energy budget of the Universe – 99% of only this 5% is in the plasma state. The rest 70% of the total is in the form of dark energy that is supposed to cause the accelerated expansion of the Universe, and 25% in the form of dark matter whose gravity keeps the galaxies from flying apart (see Fig. 1).

The standard cosmological model, which describes the evolution of the Universe as a whole, is well constrained thanks to recent missions such as Planck.[$] According to this model, the mass/energy content of the various components of the Universe (left pie-chart in Fig. 1) governs its evolution. The cosmological model and the observational determination of its handful of parameters represent a triumph of the reductionist physical approach. However, we must be mindful that we do not have a clue about 95% of the Universe. Moreover, even the budgeting of the ordinary matter (baryons) into various forms (right panel of Fig. 1) has been fairly recent and is still uncertain.[1,2] One must also remember that the state of baryons depends on the age of the Universe, which is related to the cosmological redshift defined as the ratio of the observed and the emitted wavelength of a source at a cosmological distance ($\gtrsim$ 100 Mpc, within which the matter is affected by the pull of local galaxies). Because of a finite speed of light, cosmologically distant objects show us how the universe was like at earlier epochs. So looking at astrophysical objects at large distances is like going back in time to see our younger Universe. See Fig. 2 for a sketch of the history of the Universe.

In the currently-accepted hot Big Bang cosmological model, the Universe started as a hot, dense fireball that expanded and cooled. As the Universe cooled below 3000 K, about 400,000 years after the Big Bang, the electrons and protons in the primordial plasma recombined to form neutral hydrogen. The temperature at which recombination happens is a rather simple application of Saha's ionization equation (see section 8.1 of reference 4 for a detailed discussion). The cosmic microwave background (CMB) is

---

[†] Since we use common acronyms throughout the article, we provide a glossary at the end to aid the reader. The literature cited is not at all exhaustive.

[$] https://www.esa.int/Our_Activities/Space_Science/Planck

[*] Department of Physics, Indian Institute of Science, Bangalore - 560012, e-mail : prateek@iisc.ac.in

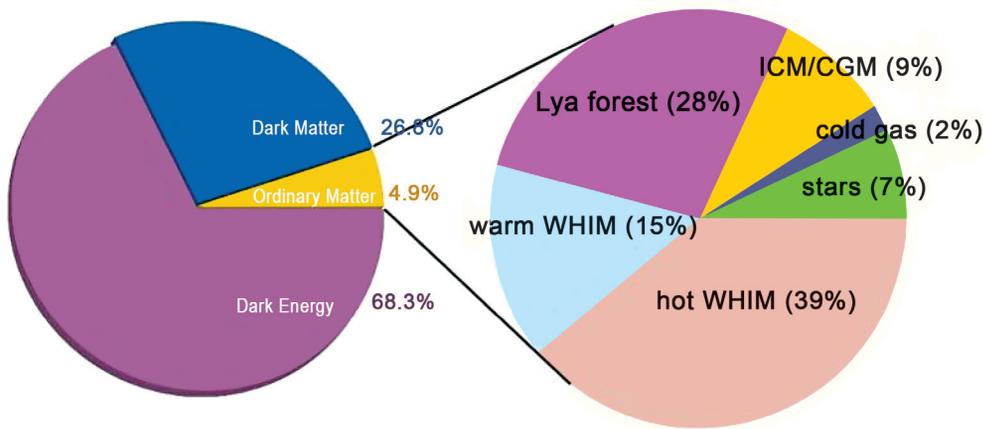

Fig. 1: The mass/energy budget of the Universe as a whole (left) and the mass budget of the ordinary matter in the local Universe (right). Most of the known matter is in the diffuse plasma state (WHIM [warm-hot intergalactic medium in cosmological filaments], warm [~$10^4$ K] IGM [intergalactic medium] traced by the Ly$\alpha$ forest, ICM/CGM [intracluster/circumgalactic medium around galaxies and clusters]). While the pie-chart on the left is fairly accurate, some of the diffuse components in the pie-chart on the right can have up to a factor of two uncertainty. (Based on references 1 and 3)

the primordial radiation coming from this epoch, after which matter becomes neutral and transparent to this black-body radiation. The next important epoch of reionization starts when the first sources of ionizing radiation - stars and accreting black holes - are born. As star formation proceeds, most of the baryons in the Universe are reionized within a billion years after the Big Bang. The ionization history of the Universe is an area of active research, but away from the main theme of this review.

The idea that matter in the Universe should concentrate in some places and rarify in others, goes as far back as Newton. Modern surveys of galaxies, and numerical simulations of dark matter particles interacting only via gravity (their non-gravitational interactions among themselves and with baryons are deduced to be weak) show a large scale structure in the Universe in form of filaments, halos and voids (Fig. 3). The cosmological filaments and sheets are not fully relaxed; they are still collapsing gravitationally. On the other hand, halos moulded by gravitational collapse at the intersection of filaments are almost spherical and relaxed, with roughly isotropic distribution of dark matter particle velocities. The diffuse plasma in cosmological filaments, called the intergalactic medium (IGM), has a broad temperature range. Most of the baryonic mass is in the temperature range of $10^5$ to $10^7$ K, called the warm-hot intergalactic medium (WHIM). The cooler phases of the IGM (~$10^4$ K) have a very small fraction of neutral hydrogen, but that is enough to produce distinct absorption features in the spectrum of a background quasar (bright point-like sources at cosmological distances) due to the Lyman-alpha transition (n=1 to 2). These absorption lines appear in the spectrum as the radiation from quasar traverses the intervening neutral hydrogen of the cooler IGM (~$10^4$ K; see Ly$\alpha$ forest marked in the right pie-chart of Fig. 1). The IGM is so dilute that it can only be detected in absorption in optical, UV, and X-rays against bright sources such as quasars.

The virialized (with kinetic energy in random isotropic motions ≈ gravitational energy) halos have somewhat denser and hotter plasma compared to the IGM. These halos are fed by the matter falling gravitationally through the filaments. The halos of different masses are expected to be the scaled version of each other, if non-gravitational processes (e.g., heating and cooling) are ignored.[5] The more massive halos, such as galaxy clusters

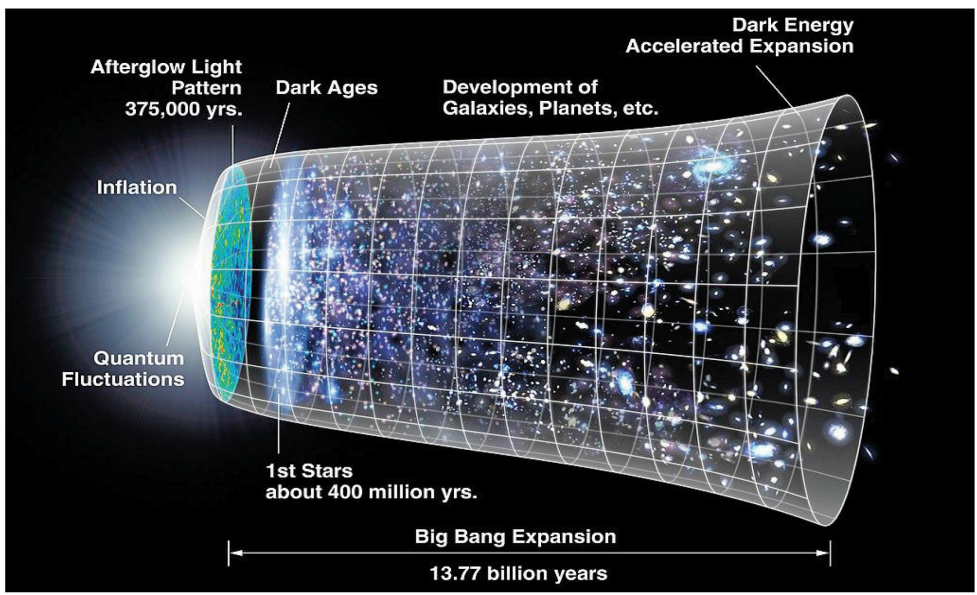

Fig. 2: The history of the Universe, with the important epochs marked. (Figure credit: Wikipedia)

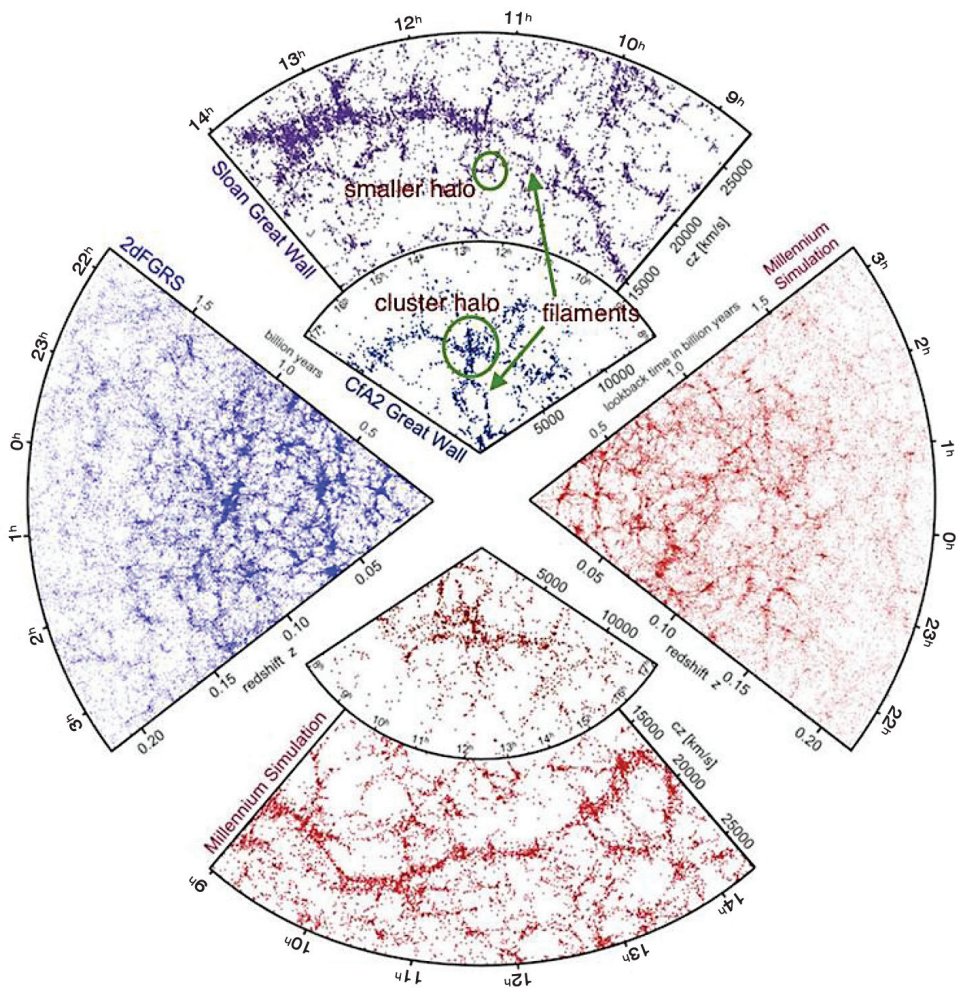

Newton. X-ray observations were combined with observations in other wavebands, such as radio, to constrain the ICM (see Fig. 5). Since the CGM is less dense and more compact than the ICM, X-ray emission is hard to detect. However, the installation of the cosmic origins spectrograph (COS) onboard the Hubble Space Telescope (HST) opened a new window in UV absorption to the CGM of nearby Milky Way-like galaxies. One of the fascinating results based on COS observations is that the multiphase gas in the CGM (ranging from $10^4$ to $3 \times 10^5$ K) has more mass than all the stars of the central galaxy![9] There is similar or even larger mass in the hot phase (~$10^6$ K) in which this cooler multiphase gas is embedded.[10] Such a Galactic "corona" confining the cooler clouds far from the Milky Way disk was proposed by Spitzer way back in 1956![11]

**Fig. 3:** Large scale structure of the Universe traced by galaxies nearby different surveys (blue & purple) and by dark matter simulations (red). Each dot in the figure indicates a galaxy. The distribution of galaxies is not uniform but forms a web-like structure. The statistical agreement between observations and simulations is striking. A cluster halo (hosting an ICM), a galactic halo (hosting a CGM), and two cosmological filaments (traced by WHIM) are marked. (Based on Reference 8)

with temperature $10^7$-$10^8$ K, have a denser plasma (known as the intracluster medium or ICM) in their cores and emit X-rays. X-ray observations show that the hot plasma in several clusters, known as relaxed or cool-core clusters, is spherically symmetric and close to hydrostatic equilibrium (just like earth's atmosphere). The lower mass halos around galaxies such as our Milky Way have a lower plasma density and temperature, and is called the circumgalactic medium (CGM). Because of its low density the CGM is very difficult to detect in emission (emissivity scales as the square of the number density for a plasma). Only our Milky Way's halo has been constrained by X-ray emission studies;[6,7] CGMs of other galaxies are simply too faint.

Because the ICM is bright in X-rays, the understanding of the ICM was revolutionized by the advent of modern X-ray telescopes such as Chandra and XMM-

### The Missing Baryons

The recent observations in UV[9] and X-rays[1] have contributed to the solution of the missing baryon problem - namely that the easily observable stars and the interstellar medium (ISM, the dense multiphase gas in form of a disk out of which new stars are born) are only a small fraction of the expected baryon abundance, both in individual halos and in the Universe as a whole. It is only now that we know the global baryon budget within a factor of two or so. The next question is to understand the state of baryons in different halos. Why are most baryons in the hot ICM for clusters but the fraction in stars increases for lower mass halos?[12] For halos less massive than our Milky Way, the fraction of baryons in stars again falls down.[13] One of the main aims of modern galaxy formation is to unravel the physics behind these observational facts (see Fig. 4).

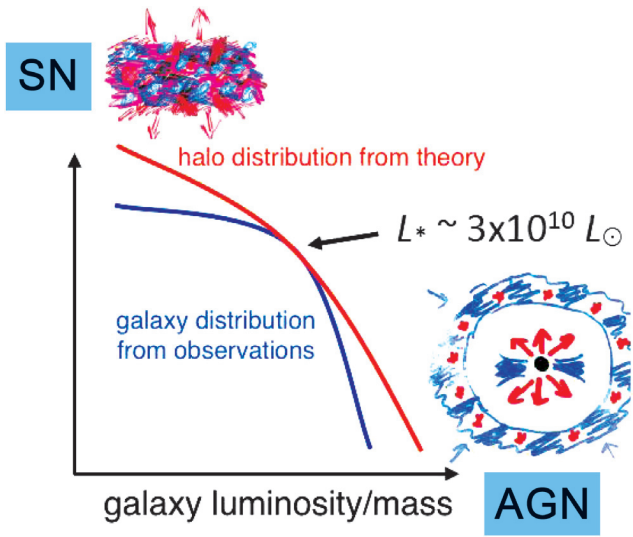

**Fig. 4:** A schematic comparison of the dark matter halo mass distribution (red line) and galaxy mass distribution (blue line). Galaxy growth is suppressed relative to the halo for both low and high halo masses because of supernova and AGN (active galactic nuclei, powered by central supermassive black holes in galaxies) feedback, respectively. (adapted from Reference 14)

### Cluster Cooling Flow Problem & Need for Feedback

One of the lessons that has emerged from recent studies is that energy injection in the ICM/CGM is crucial to explain the galaxy mass-halo mass relation. Fig. 4 shows that the conversion of baryons into stars is most efficient for Milky Way-like intermediate mass halos. The stellar fraction within halos decreases for both smaller and larger halos. It is believed that supernova explosions due to the death of massive stars can provide enough energy in the CGM to prevent it from forming stars efficiently. For deeper potential wells of galaxy clusters supernova feedback is ineffective, and radio jets powered by accretion onto the central supermassive black holes (SMBHs) are required to prevent the ICM from runaway cooling (see Fig. 5).

In absence of AGN (active galactic nuclei powered by SMBH accretion) jet heating, the dense ICM core (~0.1 cm$^{-3}$; of course, this is negligibly small compared to terrestrial densities!) can cool in just a few hundred million years, much less than the age of the cluster. Therefore, had there been no feedback, the star formation rate and mass of cold gas in cluster cores would have been much larger than what is observed. Moreover, X-ray spectra lack soft X-ray lines corresponding to the plasma cooling below $10^7$ K.[15] There is a dramatic lack of very massive blue galaxies than what the cooling-only models predict.[16] All evidence points to thermal balance in cluster cores rather than runaway cooling. The radiative losses are compensated by mechanical energy injection due to AGN jets[17] (Fig. 5).

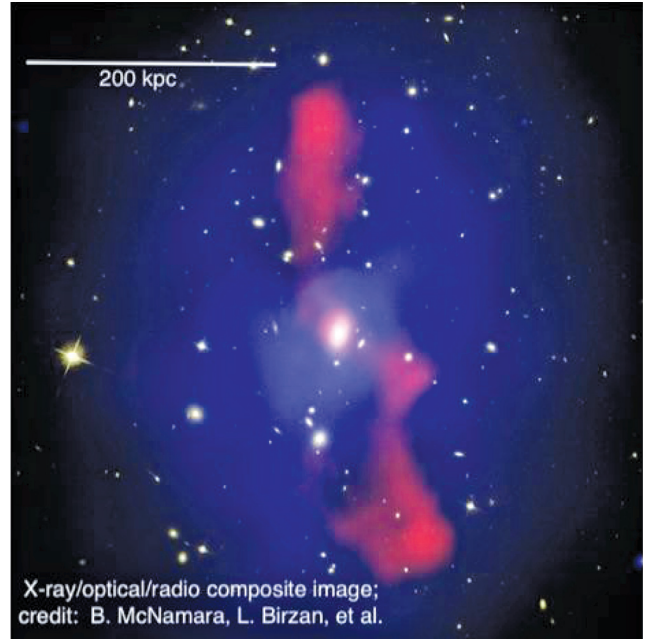

**Fig. 5:** This composite image (X-rays in blue and radio in magenta) of galaxy cluster MS9735.6+7421 shows X-ray cavities (depressions in X-ray brightness) produced by powerful AGN radio bubbles which displace the ICM plasma. Such cavities are fairly common in cool core clusters but typically much smaller (10s of kpc). The AGN jets inject mechanical energy in the ICM that dissipates and prevents catastrophic cooling losses.

### Thermal Instability and Precipitation

The focus of the article till now has been observational. A natural question is whether we can understand the state of baryons in dark matter halos from general physical principles. In this section we describe a useful model to understand the state of baryons within halos to which our group has contributed.

We can understand the ICM if we consider *local* thermal stability of a hydrostatic atmosphere in *global* thermal balance. The assumption of average thermal balance in cluster cores is motivated by observations. The ICM plasma emits X-ray photons via free-free emission — electrons accelerated in the electrostatic field of ions emit radiation. These photons are lost as soon as they are born (causing radiative losses) because the ICM is very dilute and hence transparent (for comparison, a photon produced in the core of the opaque sun takes ~30,000 years to emerge from the surface!). The emissivity is proportional to the square of particle number density ($n^2$), proportional to both the number of electrons (intensity scales with the number of electrons) and the number of protons (a higher scattering rate is expected for a higher density of ions).

It is useful to define a cooling time as the ratio of the thermal energy density ($e$) and the cooling rate density $n^2 \Lambda[T]$ ($\Lambda[T]$ is a temperature-dependent cooling function that incorporates the radiation microphysics),

$$t_{cool} = \frac{3}{2} \frac{n k_B T}{n^2 \Lambda} \propto \frac{T}{n\Lambda},$$

which is shorter for a higher density plasma. For free-free emission, $\Lambda \propto T^{1/2}$. The ICM plasma is locally thermally unstable.[18] Imagine a slightly dense region in a uniform plasma in thermal balance (heating balances cooling globally). It has a shorter cooling time $(t_{cool} \propto 1/n)$ than its surroundings and hence it cools. Since the cooling time is long compared to the time for sound waves to propagate across the volume, a roughly constant pressure is maintained. The cooling region is denser (because of pressure balance) and hence cools even faster $(t_{cool} \propto 1/n)$, resulting in local thermal instability. Cooling stops only when the plasma cools below a temperature at which radiation cannot be emitted efficiently. The stable phase occurs at ~$10^4$ K, below which atomic lines that emit efficiently cannot be excited. Thus, the dense ICM is expected to have cold gas (~$10^4$ K) condense out of the hot ($10^7$-$10^8$ K) ICM.[19,20]

Cooling and heating are not the only players in the ICM. The gravity due to the dark matter halo plays a very crucial role. Beginning in 2011, our group has discovered that the ratio of the cooling time ($t_{cool}$) to the gravitational free-fall time ($t_{ff} = [2r/g]^{1/2}$; the time that a mass will take to fall from a radius $r$ to the center of the cluster) determines if cold gas can condense out of the hot ICM.[21,22] A similar criterion applied over the whole halo (rather than at each radius) was used earlier to explain the range of observed galaxy luminosities (blue curve in Fig. 4).[23]

From idealized numerical simulations of the ICM, with heating balancing cooling at each radius (but not at every point in the shell), our group discovered that multiphase gas condenses only when $t_{cool}/t_{ff}$ is smaller than roughly 10. It is easy to understand this qualitatively. A blob slightly denser than its surroundings sinks due to buoyancy, and its motion relative to the background causes it to get mixed before it can cool to the stable temperature ($10^4$ K) if the free-fall time is short compared to this threshold. In the other regime ($t_{cool}/t_{ff} < 10$) thermal instability leads to multiphase condensation before gravity is able to shear away the cooling dense blob. This threshold of 10 is still not understood quantitatively as the physics involved is highly nonlinear.[24] However, we believe that all subsonic "coronae," not just the ICM, should have an upper limit on hot gas density corresponding to $t_{cool}/t_{ff} \gtrsim 10$ (recall that $t_{cool} \propto 1/n$). If the density is larger than this threshold, cold gas should condense out of the hot phase. In fact, $t_{cool}/t_{ff} \sim 10$ in the lower solar corona, even though the cooling and free-fall times are 10 orders of magnitude shorter than cluster cores![25]

### Feedback Regulation by Cold Gas

Once cold gas condenses out, being heavier than the hot ICM, it falls toward the center and powers the central SMBH[26] (also star formation and supernovae). Accretion onto the SMBH results in large power being put out in the ICM in form of AGN jets and bubbles, which compensate cooling in the core. This is a negative feedback loop, in the sense that larger cooling leads to more cold gas condensing and accreting on to the central SMBH that produces larger heating due to AGN jets.

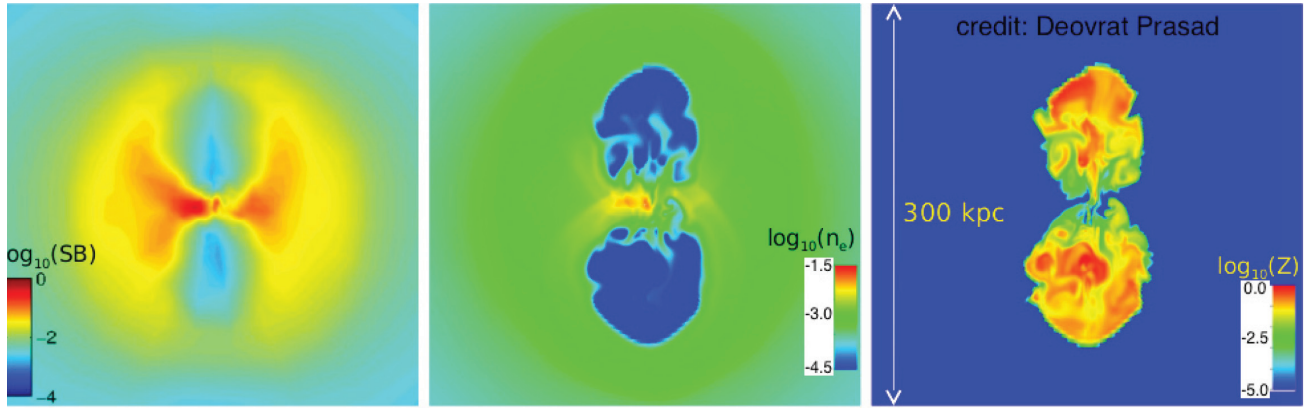

**Fig. 6:** Various diagnostics from an ICM simulation [based on the NFW+BCG run in the article of D. Prasad, P. Sharma and A. Babul, AGN jet-driven stochastic cold accretion in cluster cores, Monthly Notices of Royal Astronomical Society, 471, 1531 (2017)] after a feedback event: X-ray surface brightness map (left panel; emissivity integrated along line of sight), density snapshot in the mid-plane (middle panel), and the snapshot of the tracer of jet material (right panel). The low density buoyant bubbles appear as depressions in the SB map (consistent with the radio/X-ray observations of cool core clusters; see Fig. 5).

Beginning in 2012, simulations of the ICM interacting with feedback AGN jets over cosmological timescales have matured enough to produce results that (at least qualitatively) match observations (see Fig. 6).[27,28,29] These simulations agree with the findings of the thermal instability model and the threshold for the presence of extended cold gas. Moreover, these simulations show that the cluster core is not in perfect global thermal balance. Instead, there are cooling and heating cycles, triggered by multiphase cooling and enhanced AGN feedback in the cooling phase, and overheating of the core and suppression of further cooling in the heating phase.

Not only simulations, observations too seem to agree with the $t_{cool}/t_{ff} \approx 10$ threshold for cold gas precipitation, enhanced star formation, and strong AGN feedback.[30,31,32] One observes cool Hα filaments (tracing the $10^4$ K gas), recent star formation and powerful radio jets whenever the cluster core falls below this threshold. The feedback jets can drive turbulence in the core, but the measured turbulent velocities in the core of Perseus cluster are smaller[33] than what is required for turbulent heating to balance radiative losses in the core.[34] Thus, a heating mechanism that does not generate large turbulent velocities (e.g., turbulent mixing, shocks, thermal conduction) is preferred. Incidentally, the turbulent velocity in Perseus was measured by modeling the X-ray emission lines of highly ionized iron by the Japanese Hitomi satellite which had an unprecedented spectral resolution. Unfortunately this satellite was lost after about a month of its launch. Efforts are underway for follow-up missions that can measure the turbulent velocity in a large sample of clusters.

### *Circumgalactic Medium*

The hot ICM, being denser, is easily observable in X-ray emission. On the other hand, lower mass halos (our Milky Way resides in a $10^{12}$ solar mass halo, much smaller than cluster halos $\sim 10^{14-15}$ solar masses) have much lower density in the CGM (this can be understood from our $t_{cool}/t_{ff}$ model![35]). Therefore emission (which scales as the square of the plasma number density) is very difficult to detect† and most constraints come from recent UV absorption studies (see Fig. 7). Recall that most of the important atomic transitions of common elements, such as the Lyman-α transition (1216 Å, corresponding to n=1 to n=2) in neutral hydrogen, occur in UV.

Cooling is more efficient in Milky Way mass halos because the temperature is low enough ($\sim 10^6$ K instead of

---

† The Milky Way CGM has been detected both in emission and absorption.[10,36]

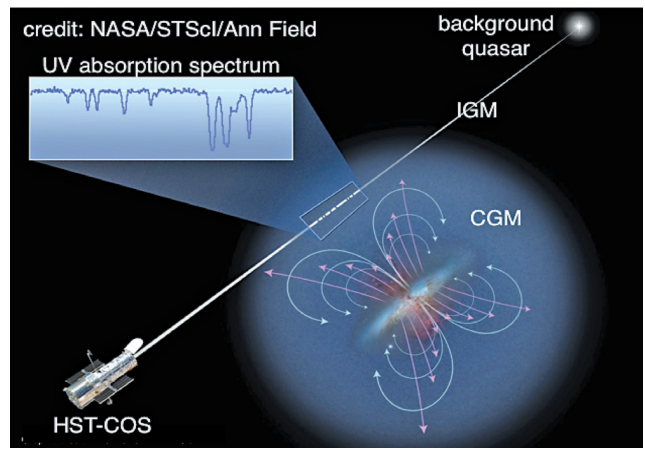

**Fig. 7:** An artist's impression of the CGM and its COS observation. CGM is the diffuse multiphase gas sphere (~200 kpc) surrounding the central galaxy that may drive outflows (shown by arrows). HST's line of sight to a background quasar intersects part of the CGM, which imprints absorption features in the quasar continuum that give clues about the properties of the CGM.

$\geq 10^7$K as in clusters) that the atoms are not fully ionized, and the transition of electrons from a higher energy level to a lower one can efficiently cool the plasma. Thus our $t_{cool}/t_{ff} \approx 10$ threshold implies a lower plasma density in lower mass halos ($t_{ff}$ is expected to be similar across different halo masses), something that can help explain the halo missing baryon problem and other properties of lower mass halos.[35] The plasma density profiles deduced from our model applied to the Milky Way halo is consistent with the observational constraints.[36,37,38] The CGM in low mass

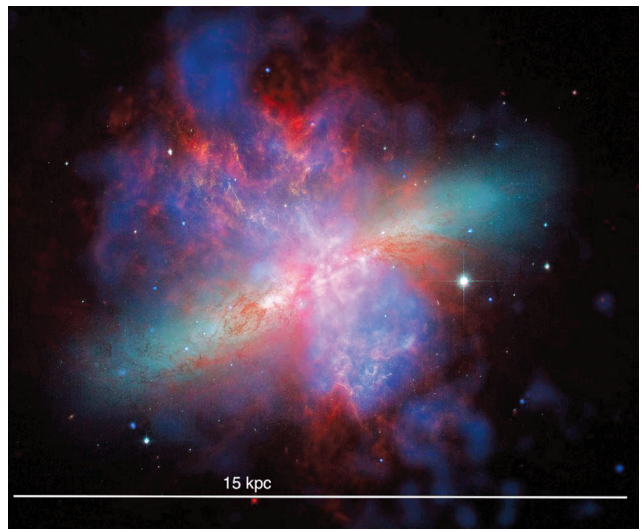

**Fig. 8:** A composite image of Messier 82 (M82), the poster-child of starburst-driven galactic outflows. Blue: X-rays from Chandra, green: visible light from HST, and red: IR from Spitzer. The multiphase outflow is perpendicular to the stellar disk (~15 kpc across). X-rays are produced by stellar wind shocks and supernovae, IR is produced due to stellar light reprocessed by dust, and visible light traces the stars in the disk. (Figure credit: Wikipedia)

halos, being less tightly bound gravitationally than the ICM, is affected much more by supernova/AGN feedback and is expected to have a larger turbulent support and higher density perturbations.[39] So the thermal instability models may need to be tweaked accordingly to apply to the CGM observations.

HST-COS can detect absorption features of several ions, present from $3 \times 10^5$ K (OVI, five times ionized oxygen ion[40]) to $10^4$ K (SiII, HI[41]) tracing a variety of temperatures/densities, in the background light of bright quasars. The absorption line can be modeled to give the column density (the integral of number density of a particular ion along the line of sight; $N_i = \int n_i dl$), and thermal and turbulent line broadening along the line of sight. Going from these observables to physical conditions (number density, temperature, etc.) is non-trivial because the turbulent plasma will have a range of densities, temperatures and velocities along the line of sight, and typically the modeling of absorption lines assumes a uniform absorber. Moreover, ions can be excited both by collisions (collisional ionization) and by radiation (photoionization), and the ionization correction (fraction of an element in a given ionization state) depends on it. The uncertainty in the structure and distribution of clouds can give more than an order of magnitude variation in the derived mass fraction of the cold gas.[9,42,43,44] In fact, the recent interesting work of McCourt et al.[45] suggests that the cold gas can fragment into cloudlets of tiny (~pc) size due to fast cooling, and these can cover a large area with a very small mass fraction, just like terrestrial mist on cold days.

Traditional galaxy formation simulations have focussed on the dense components within halos, cold gas and stars. Standard smooth particle hydrodynamics (SPH) codes, using which the first galaxy formation simulations were carried out,[46] automatically give a higher resolution in dense regions; for constant mass SPH particles the spatial resolution is ~10 times higher in the ISM compared to the diffuse CGM ($\Delta x \propto \rho^{1/3}$). Moreover, the SPH method suppresses Kelvin-Helmholtz and Rayleigh-Taylor instabilities that shred dense gas moving relative to the hot diffuse medium.[47] Eulerian simulations, that use a fixed volume of the grid cells, also suffer from other numerical problems such as the violation of Galilean invariance at unresolved interfaces.[48,49] Therefore, careful idealized numerical simulations are necessary to model the multiphase CGM spanning several decades in density/temperature and velocities.[50,51] Simply relying on cosmological simulations with necessarily inadequate resolution in the CGM will very likely give misleading results. It is no surprise that different cosmological simulations, using different numerical methods but similar feedback and cooling physics, have disagreements on basic questions such as the fraction of cold, warm and hot gas in the CGM (e.g., see Fig. 8 in reference 42). Understanding the physics of multiphase, magnetized and conducting CGM via idealized "wind tunnel" numerical experiments (analogous to the wind tunnel experiments done for aircraft design) is necessary for gaining insight into important CGM processes.[52,53]

## *Multiphase Outflows*

Multiphase galactic outflows are observed in their full glory in several nearby galaxies (see reference 54 for a review). These outflows are crucial for transporting metals into the IGM, which are observed in high redshift IGM absorption studies.[55] Outflows (traced by the flow of multiphase gas away from the galactic center) also play a fundamental role as a feedback agent over all redshifts;[56] they are necessary to explain the stellar mass-halo mass relation (Fig. 4). Outflows can be driven both by star formation (e.g., galactic outflow in M82; see Fig. 8) and by AGN (Fig. 5).

Our own Milky-Way also shows signatures of multiphase outflows.[57,58,59,62] This was dramatically highlighted by two ~55⁰ gamma-ray bubbles perpendicular to the Galactic plane detected by FERMI.[58] Such bubbles, a signature of galactic-scale feedback, should be common in most galaxies several times over their lifetimes. It is unclear if the Fermi Bubbles are powered by supernovae or by accretion on to the SMBH at the center of our Milky-Way, but X-ray observations (OVIII/OVII emission line ratios) can shed light on this.[61]

While nearby galactic outflows such as M82 are observed to be multiphase over ~10 kpc scales, a key unresolved question is how far can such a multiphase gas survive? The hot outflow is driven by overlapping supernovae and shocked stellar winds. The hot wind can lead to condensation of cold gas via thermal instability[62] or it can entrain the cooler ISM material.[63] The Kelvin-Helmholtz and Rayleigh-Taylor instabilities at the interface of cold and hot phases can shred and mix the cold clouds before they can be accelerated significantly by the ram pressure (pressure experienced by clouds due to the headwind) of the hot outflow (ref. 51; but see ref. 50 for a counterpoint). Another point is that the multiphase galactic outflow is interacting with the CGM, which can stop and mix it at ~10s of kpc rather than allowing it to propagate out to ~100 kpc.[64] Therefore, it appears that multiphase outflows may not be able to explain the ubiquitous cold gas extending out to ~100 kpc, observed isotropically around Milky Way-like galaxies observed by COS. A more natural source of multiphase gas is thermal instability in the CGM, as it is in cool core clusters.

### *Conclusions*

Although we mention the large scale intergalactic medium (IGM), our focus in this article has been the diffuse gas in and around galaxy halos (CGM/ICM). Since we could observe only stars until very recently, our understanding of galaxies is biased towards explaining only the stellar properties. We now know that stars form a small fraction of baryons across all halo masses. Moreover, stellar properties (age, metallicity distribution, etc.) suggest the fundamental importance of continuous gas supply from the CGM and the pristine (not enriched by metals) IGM for star formation. X-ray observations from the dense ICM of galaxy clusters led the way in our understanding of the CGM, but there are several differences from massive halos such as a much larger radial extent of multiphase gas and a bigger deviation from hydrostatic equilibrium in the latter.

Because of a large range of densities and temperatures in the CGM, the resolution required for directly simulating the multiphase CGM is not achievable even for the largest available computers. Therefore, physical insights from high resolution idealized simulations have to be incorporated as subgrid models in cosmological simulations.

An independent probe of the CGM/ICM is the Sunyaev-Zeldovich (SZ) distortion of the background CMB due to the inverse-Compton scattering by hot ICM/CGM electrons. This, being based on the scattering of the background CMB, is very effective at detecting plasma in halos at higher redshifts because the SZ signal is independent of the redshift.[65,66] Unlike absorption or scattering, the signal due to emission decreases inversely with the square of the distance. Highly sensitive successors of Chandra X-ray telescope, such as Lynx and Athena, shall have the sensitivity to detect the dilute hot plasma in the CGM in emission. Only by combining multi-wavelength observations and high resolution simulations can we achieve a satisfactory understanding of the CGM.

We can only cover a small fraction of exciting research going on in the area of diffuse plasmas around galaxies. Readers interested in knowing more about the CGM and galactic outflows should refer to the webpage of a conference that we organized recently covering the topics of this review.† In this article we have not talked about the role of magnetic fields and cosmic rays (the non-thermal components) in ICM/CGM. Diffuse radio emission on the scale of the whole cluster is fairly common and implies the existence of magnetic fields and relativistic electrons (the origin of which is not understood) gyrating around them very close to the speed of light.[67] Similarly, the Faraday rotation of linearly polarized light passing through galaxy clusters implies the existence of turbulent micro-Gauss magnetic fields.[68] This can be produced by the stretching and overturning of magnetic fields by the turbulent ICM.

### *Acknowledgements*

We wish to thank Alankar Dutta, Smita Mathur and Jess Werk for helpful comments and encouragement.

### *Glossary*

**AGN:** Active Galactic Nuclei are centers of galaxies, emitting from radio to gamma rays, powered by accretion onto central supermassive black holes.

**CGM:** Circumgalactic Medium, the diffuse plasma in Milky Way-like (much lower mass than clusters) halos, most easily observed in absorption.

**COS:** Cosmic Origins Spectrograph is a powerful UV spectrometer installed on the Hubble Space Telescope (HST, NASA's iconic space telescope) during the 2009 servicing mission.

**ICM:** Intracluster Medium, the hot ($10^7$-$10^8$ K) and dense (0.001-0.1 cm$^{-3}$) plasma pervading the clusters of galaxies, most easily observed in X-ray emission.

**IGM:** Intergalactic Medium, the diffuse plasma with a range of temperatures tracing the cosmological filaments connecting to dark matter halos.

**ISM:** Interstellar Medium, the dense multiphase gas distributed in form of a disk confined close to the galactic stellar disk. New stars form out of the coolest gas in the ISM.

**Lyα forest:** Lyman-alpha forest is the diffuse neutral gas tracing the IGM, and which appears as a 'forest' of absorption lines in the spectra of background quasar.

**Quasars:** Quasars are AGN that are very bright in the optical, with luminosity comparable to an entire galaxy. Quasars are used as a probe of diffuse

---

† http://www.rri.res.in/bubble/bubble.html

matter in the Universe that produces absorption features in their spectrum.

**SMBH:** Supermassive Black Holes with mass $10^5$-$10^{10}$ solar mass found at the centers of most galaxies.

**WHIM:** Warm-Hot Intergalactic Medium is the diffuse plasma ($10^5$-$10^7$ K) tracing the cosmological filaments. ❐